\documentclass{sigchi}

\toappear{\scriptsize Permission to make digital or hard copies of all or part of this work for personal or classroom use is granted without fee provided that copies are not made or distributed for profit or commercial advantage and that copies bear this notice and the full citation on the first page. Copyrights for components of this work owned by others than ACM must be honored. Abstracting with credit is permitted. To copy otherwise, or republish, to post on servers or to redistribute to lists, requires prior specific permission and/or a fee. Request permissions from permissions@acm.org. \\
{\emph{CHI '20, April 25--30, 2020, Honolulu, HI, USA.} } \\
Copyright is held by the owner/author(s). Publication rights licensed to ACM. \\
ACM ISBN 978-1-4503-6708-0/20/04\ ...\$15.00.\\
\url{http://dx.doi.org/10.1145/3313831.3376262} }

\usepackage{balance}       % to better equalize the last page
\usepackage{graphics}      % for EPS, load graphicx instead 
\usepackage[T1]{fontenc}   % for umlauts and other diaeresis
\usepackage{txfonts}
\usepackage{mathptmx}
\usepackage[pdflang={en-US},pdftex]{hyperref}
\usepackage{color}
\usepackage{booktabs}
\usepackage{textcomp}
\usepackage{amsmath}
\usepackage{float}
\usepackage{listings}
\usepackage{cite}
\usepackage[utf8]{inputenc}
\newcommand\yc[1]{\textcolor{bcolor4}{#1}}

\definecolor{bcolor}{rgb}     {1.0,0.0,0.0}
\definecolor{bcolor2}{rgb}     {1.0,0.0,1.0}
\definecolor{bcolor3}{rgb}     {0.0,0.0,1.0}
\definecolor{bcolor4}{rgb}     {0.0,0.0,0.0}

\usepackage{microtype}        % Improved Tracking and Kerning
\usepackage{ccicons}          % Cite your images correctly!

\usepackage{todonotes}

\def\plaintitle{Button Simulation and Design via FDVV Models}

\def\emptyauthor{}
\def\plainkeywords{Button; haptic; modeling; simulation; tactility; force feedback; vibration; input device; haptic rendering; FD model; FDVV model.}

\newcommand{\compresslist}
{
    \setlength{\itemsep}{1pt}
    \setlength{\parskip}{0pt}
    \setlength{\parsep}{0pt}
}

\makeatletter
\def\url@leostyle{%
  \@ifundefined{selectfont}{
    \def\UrlFont{\sf}
  }{
    \def\UrlFont{\small\bf\ttfamily}
  }}
\makeatother
\def\pprw{8.5in}
\def\pprh{11in}

\definecolor{linkColor}{RGB}{6,125,233}
\hypersetup{%
  pdftitle={\plaintitle},
  pdfauthor={\emptyauthor},
  pdfkeywords={\plainkeywords},
  pdfdisplaydoctitle=true, % For Accessibility
  bookmarksnumbered,
  pdfstartview={FitH},
  colorlinks,
  citecolor=black,
  filecolor=black,
  linkcolor=black,
  urlcolor=linkColor,
  breaklinks=true,
  hypertexnames=false
}

\begin{document}

\title{\plaintitle}

\author{
 \alignauthor{Yi-Chi Liao\textsuperscript{1} \hspace{1cm} Sunjun Kim\textsuperscript{1,2,3} \hspace{1cm} Byungjoo Lee\textsuperscript{2} \hspace{1cm} Antti Oulasvirta\textsuperscript{1}\\
 \affaddr{\textsuperscript{1}Aalto University, Finland \hspace{0.5cm}\textsuperscript{2} KAIST, Republic of Korea}
 \hspace{0.5cm}\textsuperscript{3}DGIST, Republic of Korea
 \\
 \email{yi-chi.liao@aalto.fi},
\email{sunjun.kim@aalto.fi},
\email{byungjoo.lee@kaist.ac.kr},
\email{antti.oulasvirta@aalto.fi}
 }
}

\maketitle

\begin{abstract}
Designing a push-button with desired sensation and performance is challenging because the mechanical construction must have the right response characteristics.
Physical simulation of a button's force--displacement (FD) response has been studied to facilitate prototyping; 
however, the simulations' scope and realism have been limited.
In this paper, we extend FD modeling to include vibration (V) and velocity-dependence characteristics (V).
The resulting FDVV models better capture tactility characteristics of buttons, including snap.
They increase the range of simulated buttons and the perceived realism relative to FD models.
The paper also demonstrates methods for obtaining these models, editing them, and simulating accordingly.
This end-to-end approach enables the analysis, prototyping, and optimization of buttons, and supports exploring designs that would be hard to implement mechanically.
\end{abstract}

%\category{H.5.2.}{Information Interfaces and Presentation
%  (e.g. HCI)}{User Interfaces} \category{Haptic I/O and Input devices \& strategies.}{}{}





\begin{figure}[t!]
\centering
  \includegraphics[width=0.99\columnwidth]{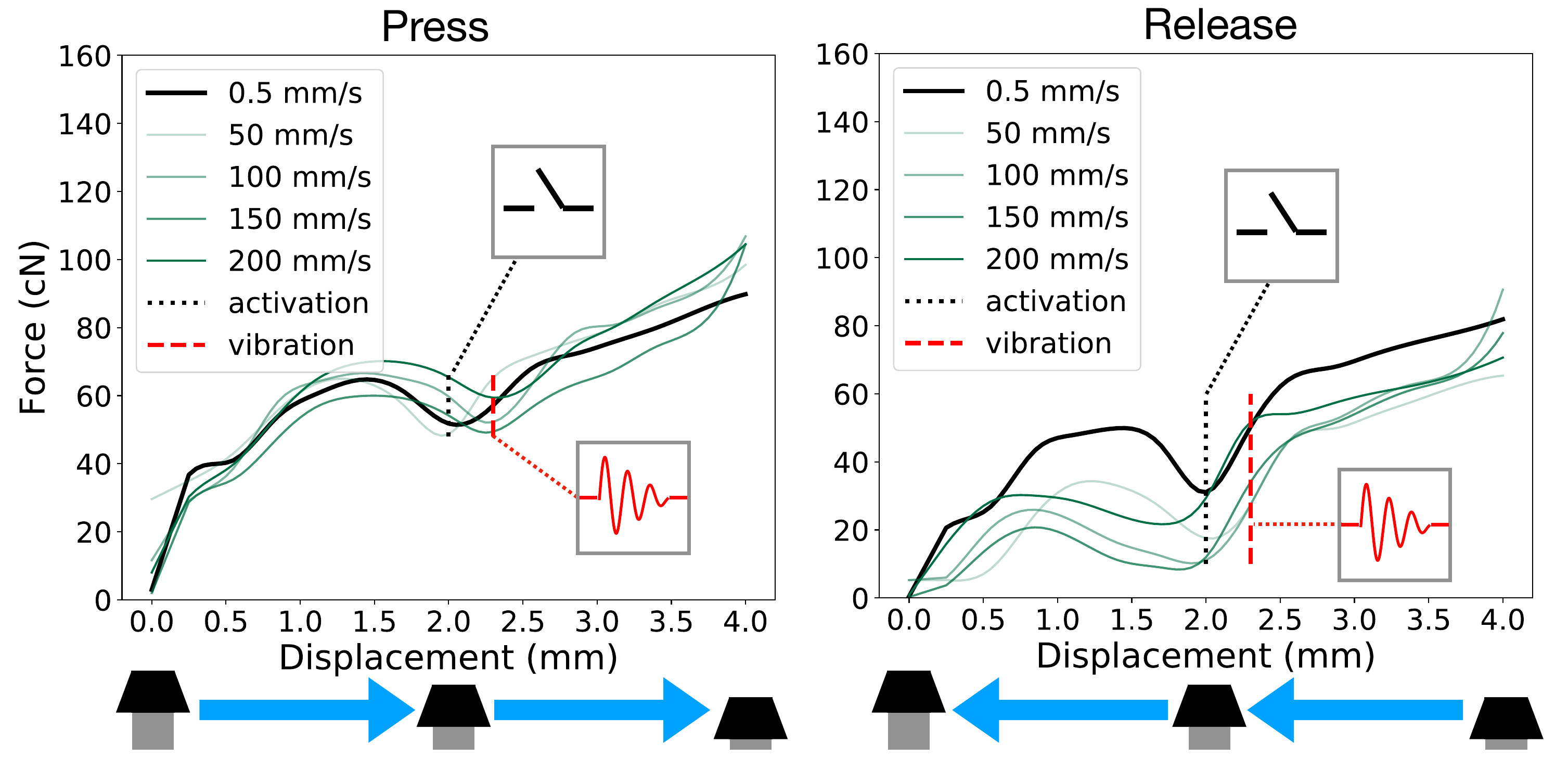}
  \caption{A force--displacement--vibration--velocity (FDVV) model represents speed-dependent physical responses of a button when pressed. 
  We show methods for capturing button presses as FDVV models, rendering them in a physical simulator, and editing and optimizing these in software. 
  The press and release models shown are for a 4 mm tactile button. Blue curves represent the corresponding (velocity-agnostic) force--displacement model \yc{typically measured by a probing machine with static and slow velocity}. 
    }~\label{fig:fdvmodel}
\end{figure}

\section{Introduction}

This paper investigates the simulation and interactive \yc{design} of push-buttons. 
Many push-button designs use a spring-loaded slider: 
when the slider is pushed to the activation point, 
a binary input is registered. Upon release, it returns to the initial state. 
More generally, buttons are transducers that register a discrete event from physical motion~\cite{impact, integration, neuro}.
Numerous types exist, using spring-loading but also other mechanisms, such as rubber and metal domes.
Interestingly, each button is unique in its \emph{tactility} or tactile response characteristics~\cite{KIM2014,velocity-fdcurve}.
Gamers, programmers, typists, and hobbyist groups alike have a keen interest in tactility,
which is associated with sensory experience and performance~\cite{keyboard,Crump2010,neuro}.
However, despite the popularity \yc{and importance }of buttons, 
researchers have paid relatively little attention to their design.

Simulators have taken on a major role in most branches of engineering and design,
where they are used to study reality, predict consequences of design decisions,
and derive and optimize solutions.
Dedicated simulators can do the same for button design.
We believe that one obstacle has been the lack of an accurate yet practical simulation device.
It should be able to realistically reproduce different tactilities
so that designers and researchers could explore and test them at little cost. 

Accurate simulation of button-pressing is challenging, though. 
Although a press and release occurs in about 100 ms~\cite{tapboard},
a wealth of sensory feedback is generated~\cite{clark_2013,neuro}.
Slow and fast mechanoreceptors ~\cite{BVallbo1999PropertiesOC, Birznieks8222,touchsense}
deliver information on spatial patterns, change in the contact area at the fingertip, the roughness of the contact support, 
stretching of the skin, subtle and rapid changes in force, and vibrations during a press.
Meanwhile, proprioception provides feedback on displacement, as detected via the joints and muscles of the finger~\cite{visual_proprioceptive}. 
This information is transmitted through the spine at high rates, up to 1~kHz~\cite{SensingSA,TactileSensing}, 
and contributes to constructing the felt sensation of a press~\cite{sensory_neuron_touch,neuro}.
Hence, to reproduce a realistic button sensation, the simulator needs to capture those physical characteristics of a press that dominate that sensation.

\begin{figure*}[ht!]
  \includegraphics[width=0.99\textwidth]{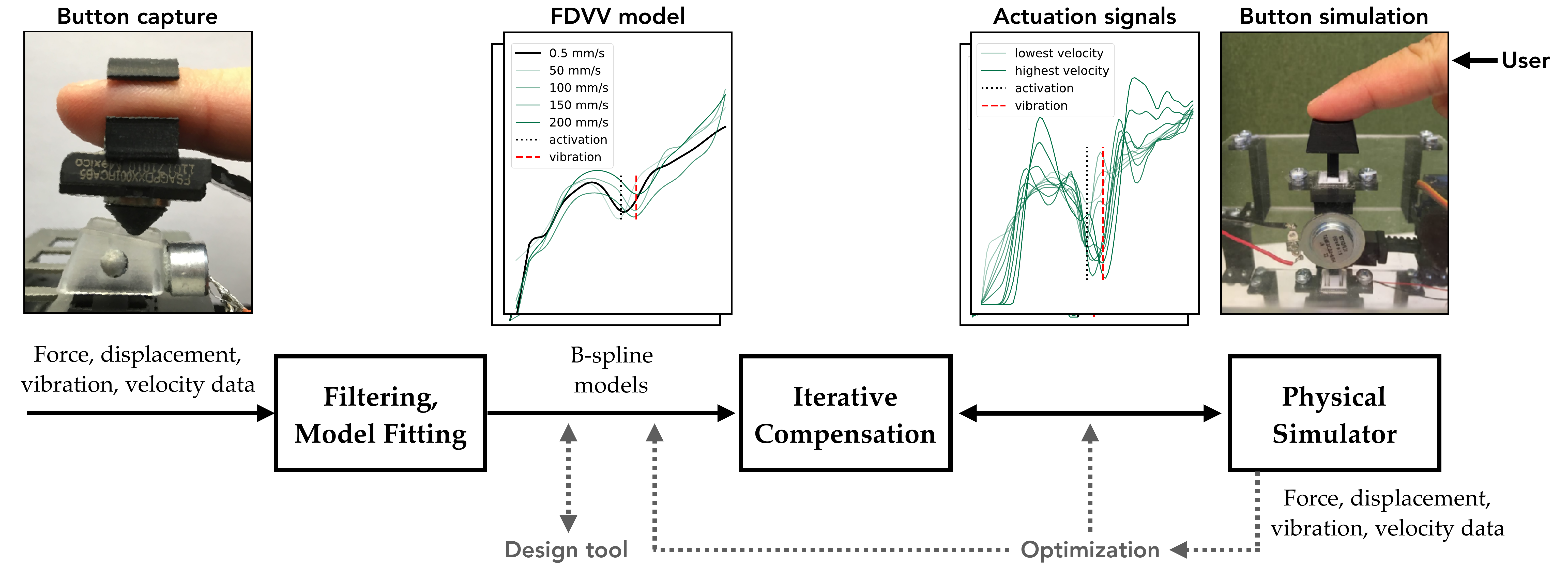}
  \caption{Overview: An end-to-end approach to button simulation. To capture an FDVV model of a button, sensors are placed on the finger, and the button is pressed multiple times. 
  The resulting force, displacement, vibration, and velocity data are filtered, and lower-parametric B-splines are fitted for use of Bayesian Information Criteria (BIC) as the fitness metric.
  A designer can edit the model produced. 
  To render the model with a given physical plant, an iterative compensation process computes how to cancel the plant's own transfer function. 
  The resulting actuation signals drive the simulator.
  }
  \label{fig:pipeline}
\end{figure*}

While pre-existing simulators can render tactile and linear buttons~\cite{Doerrer02,Liao}, we found that force--displacement (FD) approaches cannot accurately render other than a simple linear button.
In addition, the perceived realism of the rendered buttons has not been properly evaluated, 
and no methods have been offered to help designers and engineers exploit such simulation.
We believe this to be due to three engineering challenges: 
(1) modeling, (2) simulator construction, and (3) model--simulator separation.
Firstly, prior work has modeled buttons' response as the displacement-dependent change in force~\cite{Doerrer02,Liao,inforce}.
However, as we show in this paper, 
an FD model is adequate only for linear buttons pressed at extremely low speed.
It is known that the physical response of a button depends \emph{also} on velocity and acceleration~\cite{Becedas2009,AlgebraicIM,velocity-fdcurve},
and that buttons elicit a vibration response.
Vibration affects the perception of button-pressing~\cite{Ogawa2015},
and it can be so prominent that it produces an illusion of force change~\cite{psuedohaptic,3dpress,Strohmeier2017}.
Secondly, the simulators created thus far have fallen short of the operation rate needed for a button press.
General haptic devices such as Phantom~\cite{thephantom94,Phantom-based-haptic-interaction,phantom-deformable-objects} and inForce~\cite{inforce} operate at 60~Hz,
while the skin's mechanoreceptors fire at about 1,000 Hz and respond to tiny changes in force~\cite{SensingSA,TactileSensing}.
Simplistic controller approaches may have further curbed attempts to render anything other than a linear button. 
To our knowledge, previous work has, at best, applied a linear PID controller to render the response at the force actuator of a simulator, 
although this is generally recognized as insufficient for nonlinear plants~\cite{future-of-pid,Linear-Controller-Design}. 
We should emphasize, thirdly and finally, that the button models must be editable if they are to be of practical use.
This calls for model--simulator separation.
By avoiding the device-dependency of models, we can support the meaningful editing of buttons -- by designers and by software.

To address these challenges, we propose an extended model and an end-to-end simulation pipeline around it.
Our approach allows simulating more button types than previously, among them tactile-type buttons and buttons with different click reactions and travel ranges.
Furthermore, it permits the analysis and editing of buttons. 
Our work centers on the \emph{Force--Displacement--Vibration--Velocity} (FDVV) model, illustrated in Figure \ref{fig:fdvmodel}.  
It adds vibration response and velocity-dependence on top of the FD model. 
In our implementation, vibration is sampled thorough a microphone during a switch press, and multiple FD curves are sampled, at several speeds.
We solve several engineering challenges connected with ambitions to capture and simulate buttons via FDVV models. 

In particular, we present methods developed for 
(1) capturing the FDVV response of a button, 
(2) computing a lower-parametric FDVV model from the measurement data obtained, 
and (3) actuating an FDVV model for a given physical plant.
Figure \ref{fig:pipeline} gives an overview of the end-to-end approach.
The data (force, displacement, vibration, and velocities) obtained during capture are filtered and modeled via B-splines~\cite{bspline}, collapsing the data into a lower-dimensional and more manipulable model. 
For rendering it, we present a novel simulator construction for FDVV models. 
This is capable of detecting displacement to $\mu$m precision at a high sampling rate (1~kHz) 
and can produce a wide range of force (up to 4.4~N) and vibration (50~Hz~--~20~kHz) feedback.  
In contrast to previous, 60~Hz simulators, our simulator can render high-fidelity vibration arising from the rapid force change near the snap and bottom-out points during the press \cite{virtual}. 
It also has a mechanical limiter for rendering a button with various travel distances (0--6.2~mm). 
Thanks to the iterative compensation method, which translates an FDVV model into actuation signals that cancel out the simulator plant's own transfer function, one can assign button designs from software without hardware tweaks. 

In summary, this paper makes three contributions.
We present advances in modeling and simulator design that extend the range of supported button types.
Secondly, we report the results of a controlled study showing that the FDVV model yields higher perceived realism than FD modeling. 
Finally, the approach opens new possibilities in design and prototyping\yc{; especially, by reducing the effort of exploring designs.}
We demonstrate applications in interactive button-editing, software-side optimization, and prototyping of innovative button designs the mechanical structure of which would be hard to fabricate. 
\yc{The general principles we applied in the simulation pipeline can also benefit future research toward accurate haptic rendering. For example, simulating elastic materials.}

\section{Background}

Physical buttons are electromechanical devices that make or break a signal upon pressing, 
then return to the initial (re-pushable) state upon release.
There is incredible variety in the constructions that fit this definition. 
Our discussion here focuses on commonplace push-buttons of keyboards and button panels.
Mechanical keyswitches, rubber domes, and metal domes are the most typical structures.
Numerous other design parameters exist, such as physical properties of the keycap (width, slant, and key depth), the materials used (e.g., plastics), 
and system-level feedback (modalities and latencies) \cite{lewis1997keys}.
The response upon pressing can be characterized via the \emph{the force--displacement function} or force curve \cite{ActivationForceTravel,lewis1997keys}.
Actuation (press-down) and release curves often differ.
The FD curve is known to affect not only sensation but also joint kinematics~\cite{JINDRICH2004}, 
muscle activity~\cite{Kim2014DifferencesIT,ActivationForceTravel,effectofkeyboard}, 
and the user's aiming performance \cite{neuro}. 
\emph{Linear buttons} have the feel of pressing a spring; 
there is no tactile landmark or ``bump'' during press-down. 
A \emph{tactile-type} button has a so-called snap ratio, 
which determines the strength of its tactile bump.
Rubber-dome buttons utilize a snap ratio greater than 40\%.
Some tactile buttons emit an audible ``click'' sound near the snap point. 
\emph{Travel distance} is the total distance before the keycap hits the bottom,
and the distance at which the button is activated is called its \emph{activation point} \cite{impact}. 
While these features can be modeled with FD curves, we stress again that 
FD neglects velocity and vibration \yc{characteristics}. 

\subsection{Capturing and modeling physical buttons}

There are two main approaches in haptics research applicable to the modeling of buttons.
The first is an analytical one aimed at formulating equations that capture the mechanics of haptic interaction,
which in the case of buttons would involve the forces, vibrations, etc. during a press.
Since these interactions are complex, 
analytical models almost inevitably need to make simplifying assumptions.
For example, the spring--mass-damper system in buttons could be described as a lumped mass \cite{meirovitch1997principles}.
Analytical models applicable to buttons include modeling of spring--mass-damper systems~\cite{Becedas2009} and friction~\cite{computational_friction}.
While prior applications to buttons and knobs do exist~\cite{mechatronic_device}, 
the ones presented thus far are too low-parametric to capture the rich design space of push-buttons. 

The second approach is a reality-based, or data-driven, one that starts with physical measurements and constructs models based on data.
In the case of buttons, one starts by physically probing a button to measure the interaction forces between the button and the probe as displacement and even higher-order variables.
Displacement has been approached in such a manner 
outside the button domain. 
One could cite as example applications the automotive gearshift~\cite{Angerilli}, 
non-rigid materials ~\cite{inforce,RenderingSoftness}, 
and human tissue~\cite{Pai2018}. 
Displacement-only models are relatively simple. 
For buttons, this approach is insufficient. 
There have been studies examining higher-order variables, such as velocity~\cite{haptic_profile} and acceleration~\cite{identify_nonlinear,MacLean1995,MacLean1996TheHapticCA}.
This, however, complicates everything from measurement to simulation.  
Nonetheless, we followed the reality-based approach.
We captured the forces involved, displacement values, vibration, and pressing velocities, and we found a way to collapse the data to a more understandable, lower-parametric model. 

\subsection{Haptics rendering}

Our work is aligned with haptics research pursuing the creation of rich and realistic sensations~\cite{worldoftouch}.
While this area of research is too broad to review here, 
some relevant findings are worth mentioning.
Firstly, research has looked at advanced factors affecting haptic perception, such as friction, temperature, or texture \cite{softness-display}.
However, the focus has been on exploration or manipulation of objects,
which is quite different from a keypress, which occurs in around 100 ms.
The rapid compression and mechanical vibration of tissue in the fingertip are core elements of a button press.
Secondly, general-purpose haptic simulators have been produced that could be used for buttons.
The Phantom device~\cite{thephantom94,Phantom-based-haptic-interaction,phantom-deformable-objects} is a 6-DOF pen-type
general force-rendering device capable of emulating the softness of deformable objects. 
However, a low operating rate (60 Hz), excessive degrees of freedom (six instead of one), 
and lack of vibrotactile simulation limit its use for buttons. 
Softness displays~\cite{moy2000,Song2008SoftnessHD,inforce} too might aid in simulating the stiffness of a button,
but these are restricted to so-called simple stiffness, which is inadequate for buttons. 
Finally, pseudo-force-feedback has been explored.
By changing the contact area of the finger~\cite{softness-display,Ikeda2004} or using electro-tactile displays, one can create a softness-like sensation~\cite{Kajimoto2001,Takei2015}. 
However, this is not central to commodity push-buttons' design.

\subsection{Button simulators}  

Prior work on button simulation has been limited to static (speed-agnostic) FD simulators,
which have the limitations described above.
Doerrer and Werthschuetzky~\cite{Doerrer02} enabled users to edit FD curves in software, and 
Liao et al. \cite{Liao} have presented an FD simulator. The Phantom haptic interface \cite{thephantom94} also can render a virtual button. Yet, as other simulators do, these too need an exaggerated FD curve to render the ``snap'' feeling of a tactile button.
These papers omitted velocity and vibration dynamics from their model.
Moreover, to our knowledge, the perceived realism achievable by FD simulators has not been empirically validated thus far.

\section{Button Capture}

Most previous work has captured buttons by a single-FD curve model. 
This approach has become conventional also in the way manufacturers present buttons on datasheets. 
However, button-pressing involves complex phenomena affected by the varying stiffness and damping effects produced by mechanical design. 
The \emph{stiffness effect} means that the resisting force changes as a function of the button's displacement.
The \emph{damping effect} entails the resisting force changing with the button's velocity. 
An FD-only model captures neither the damping effect
nor the high-frequency structural vibrations of a button press.

We propose an extension to the physical measurement of the tactility characteristics of push-buttons. 
Our capture method features three novel elements:
(1) We measure presses under different velocities. 
(2) A human finger is used for pressing, as opposed to a rigid, static-velocity probing object as in earlier work. 
This allows us to better cover the response envelope people encounter in everyday button-pressing, 
via a procedure that requires no more than a few minutes to complete per button.
(3) We record vibrations, which are important for covering more advanced button types.
This necessitates a more complex measurement setup than before.

\begin{figure}[t!]
\centering
  \includegraphics[width=0.99\columnwidth]{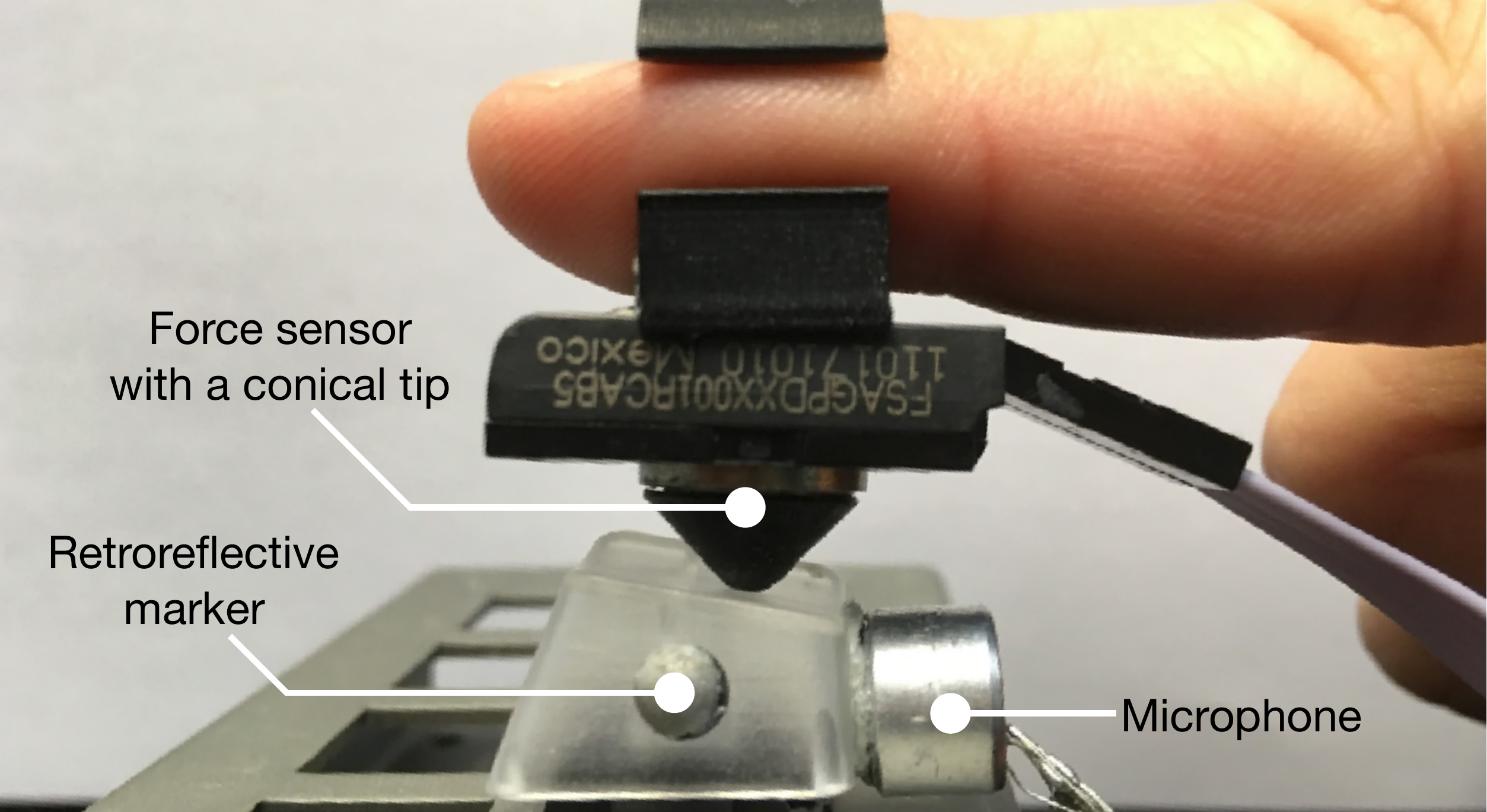}
  \caption{Button capture of a real button (4~mm tactile button). A force sensor is worn on the fingertip. Reflective markers (for motion tracking) and microphone are attached on keycap.}~\label{fig:measurement}
\end{figure}

\textbf{Setup:} Our measurement setup is shown in Figure \ref{fig:measurement}. 
Two retroreflective markers are attached to sides of the keycap. 
A motion-tracking system (OptiTrack Prime 13, 256 FPS) records the displacement of the button during a press. 
Also, a microphone \yc{(KY-038)} is attached to the keycap, to detect vibration during button presses.
On the user's index finger is a force sensor (Honeywell, FSAGPDXX001RCAB5), with a conical tip (ABS, 3D printed, \yc{Bottom and top: 12 and 5 mm-diameter circles. Height: 5 mm}) attached.
A microprocessor (Adafruit Metro M0 Express) samples both sensors with its internal 10-bit ADC at 1 kHz frequency,  
which is the highest sampling rate allowed for the force sensor.
Synchronization between the sensor data and the motion data is handled via an optical clapperboard (three 850 nm IR LEDs). 
The microprocessor is attached to a switch that turns on the LEDs. 
When the switch is triggered, the microprocessor and the motion tracking system record a reference point for synchronization. 

\textbf{Procedure:} 
A participant is asked to wear the sensors and press the given button, following the instructions on the display.
A velocity indicator is presented on this display, 
which shows animation for various velocities.
The velocity indicator also creates a beep sound indicating the moment of contact and that of hitting the bottom.
Note that we stated that the animation refers to the rate of pressing and the average velocity, 
not the moment-by-moment velocity. 
The participant is asked to press the button at the specified pace. 
We go through velocities of 50, 100, 150, and 200 mm/s,
collecting 15 presses per velocity.

\textbf{Example buttons:}
The paper reports on studies of six physical buttons:
Cherry MX Clear and Brown (4 mm, tactile), Cherry MX Black and Red (4 mm, linear), HP PR1101U (3.6 mm, tactile), and MacBook Pro 2011 (2.2 mm, tactile).
They have distinct haptic profiles (linear/tactile) and travel ranges.

\section{FDVV Modeling}

The raw measurement data must be transformed into a lower-dimensional FDVV model \yc{to allow efficient design and optimization}.
A series of preprocessing steps are followed to produce a synchronized and filtered dataset from multiple data sources. 
We then fit a B-spline model, using Bayesian Information Criteria (BIC)~\cite{Konishi2007}.  

\subsection{Preprocessing}

The outputs of button capture are  \textit{(timestamp, force, sound)} data from the microprocessor and \textit{(timestamp, the 3D position of marker1, the 3D position of marker2)} data from the motion tracker and vibration data. 
Our goals are to (1) filter out noise and outliers and (2) synchronize the two data sources.

\emph{Step 1, filtering}:
We pass force and vibration data through a low-pass filter for antialiasing, 
before the analog--digital conversion of the microprocessor. 
The filter consists of a resistor--capacitor circuit (R 333 ohm, C 1~uF) 
with a cutoff frequency of 500 Hz, using a Nyquist frequency of 1,000 Hz.
Gaussian filters ($\sigma=1.2~mm$) are then applied to both force and displacement data, separately, for further denoising. 

\emph{Step 2, synchronization}:
We then synchronize the data from the microprocessor and motion tracker, 
using the IR LED light data to find keyframes.
Because the microprocessor runs at 1,000 Hz and the motion tracker at 256 Hz, 
resampling is required before synchronizing of the two sides. 
We chose to upsample displacement data by matching the of the motion tracker to the timestamp of the microprocessor, via linear interpolation.
The displacement data get resampled up to 1,000 Hz frequency, 
after which we have synchronized the signals.
We need resampling also to register force data against displacement (here, sampled every 50 $\mu$m).

\emph{Step 3, outlier removal, averaging and smoothing}:
We use the resulting dataset to filter out incomplete button presses (ones where the button did not hit the bottom). 
Finally, \yc{we averaged the force data of the representative presses at each displacement point (every 50 $\mu$m)}.
A Gaussian filter ($\sigma=0.8~mm$) is then applied to smooth the averaged curve.

\emph{Step 4, synchronizing vibration data}:
Note that vibration data did not pass through \textit{steps 3--4}.
In most buttons, vibration is associated with the snap.
This can be programmatically exploited to synchronize the vibration signal. 
On the other hand, one can ignore measurements detected in the beginning and the end portion, 
because these are caused mostly by the finger hitting the keycap or the keycap hitting the bottom. 
Hence, we consider only the middle part of the press, 
for which we use threshold-based event detection to find the onset of the vibration. 
For some buttons, this sound wave can be very subtle and rapid (typically <25~ms), making it hard to detect programmatically.
To compensate for this, we can resort to a human observer (see ``Iterative Compensation,'' below). 

\subsection{B-Spline Fitting}

\begin{figure}
\centering
  \includegraphics[width=0.99\columnwidth]{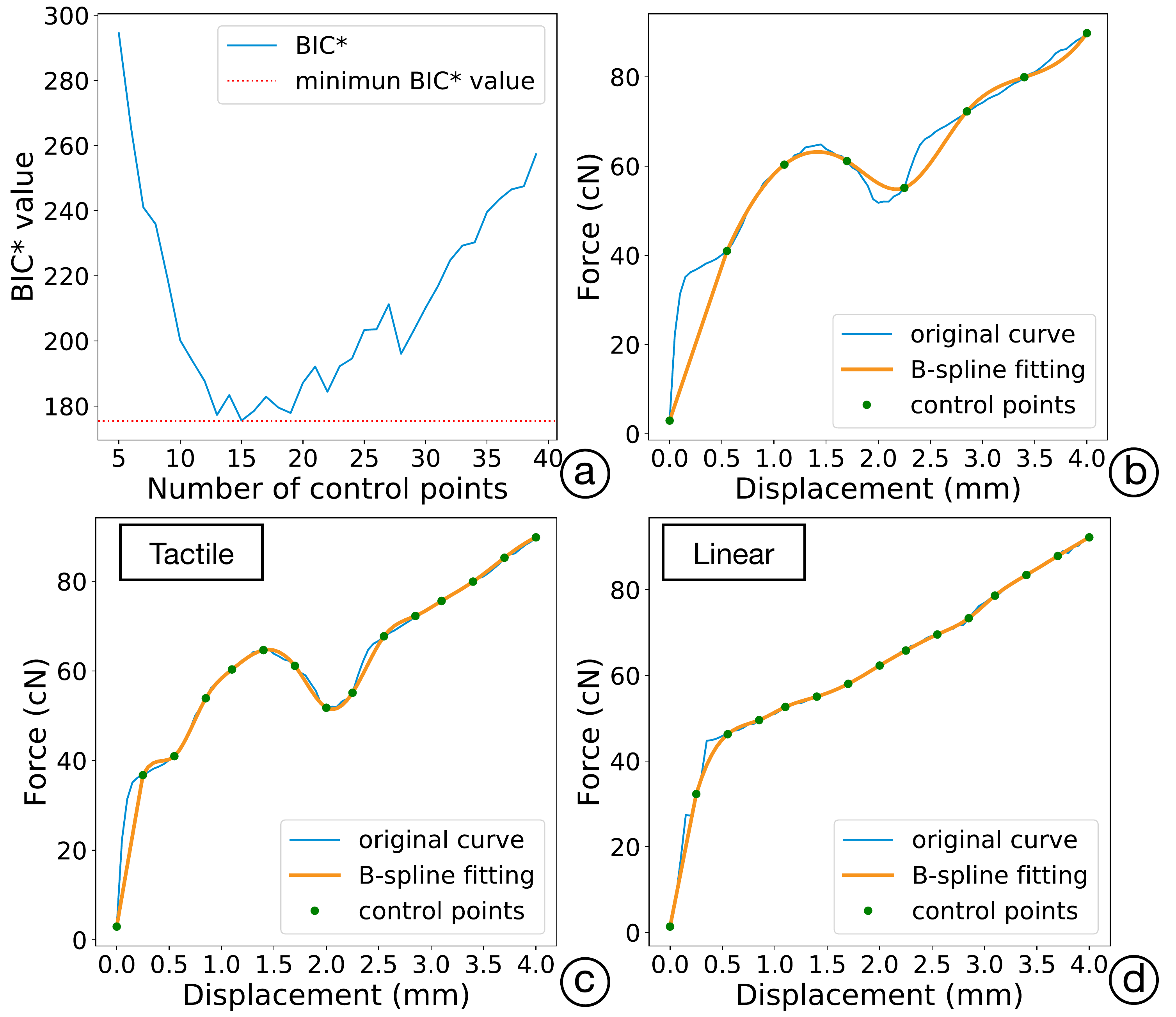}
  \caption{We use B-splines to obtain a lower-dimensional, editable FDVV model from the capture data. (a) We found that 15 control points is the ideal number of parameters to model six commodity buttons. (b) With fewer control points, the model underfits essential features of a button response. Panes c and d show example results for tactile and linear buttons (15 control points).}~\label{fig:curve_fitting} 
\end{figure}

The preprocessed dataset is still too high-dimensional for editing by designers. 
For example, a typical 4 mm button requires approximately 800 parameters in our procedure.
Hence, we use B-splines to achieve a lower-dimensional parametric model. 
While B-splines offer a suitable model for continuous multimodal data,
there is still the question of how many control points are needed. 
We studied this by fitting B-spline models to our button dataset. 
To control against overfitting, we used Bayesian Information Criterion (BIC)~\cite{bic,Konishi2007} for the fitness criteria.
To reduce the number of parameters even further for feasibility for human editing,
we added a custom penalty term: \emph{Complexity Penalty, P}. 
This results in a modified BIC\textsuperscript{*} function:
\begin{equation} \label{eq:1}
 BIC\textsuperscript{*} = \ln(n)kP - 2 \ln(\hat{L}) 
\end{equation}
Where \textit{n} stands for the number of observations, \textit{k} for the number of parameters (control points), and $\hat{L}$ for the maximized value of the likelihood function of the model. \textit{P} is the added Complexity Penalty value, which is set to 2.5. 
Figure~\ref{fig:curve_fitting} (b, c, d) gives examples of fitting for a pressing segment of a certain velocity for Cherry MX Clear button.

Following this procedure, we found 15 B-spline control points was an ideal tradeoff for the six-button dataset with root mean square error of 0.14 cN; see Figure~\ref{fig:curve_fitting} (a). 
We also registered \emph{travel range}, the \emph{activation point}, and the \emph{vibration point} in the resulting models, as examples shown in Figure \ref{fig:curve_fitting} (c, d).

\section{Button Simulator}

\begin{figure}[t!]
\centering
  \includegraphics[width=0.99\columnwidth]{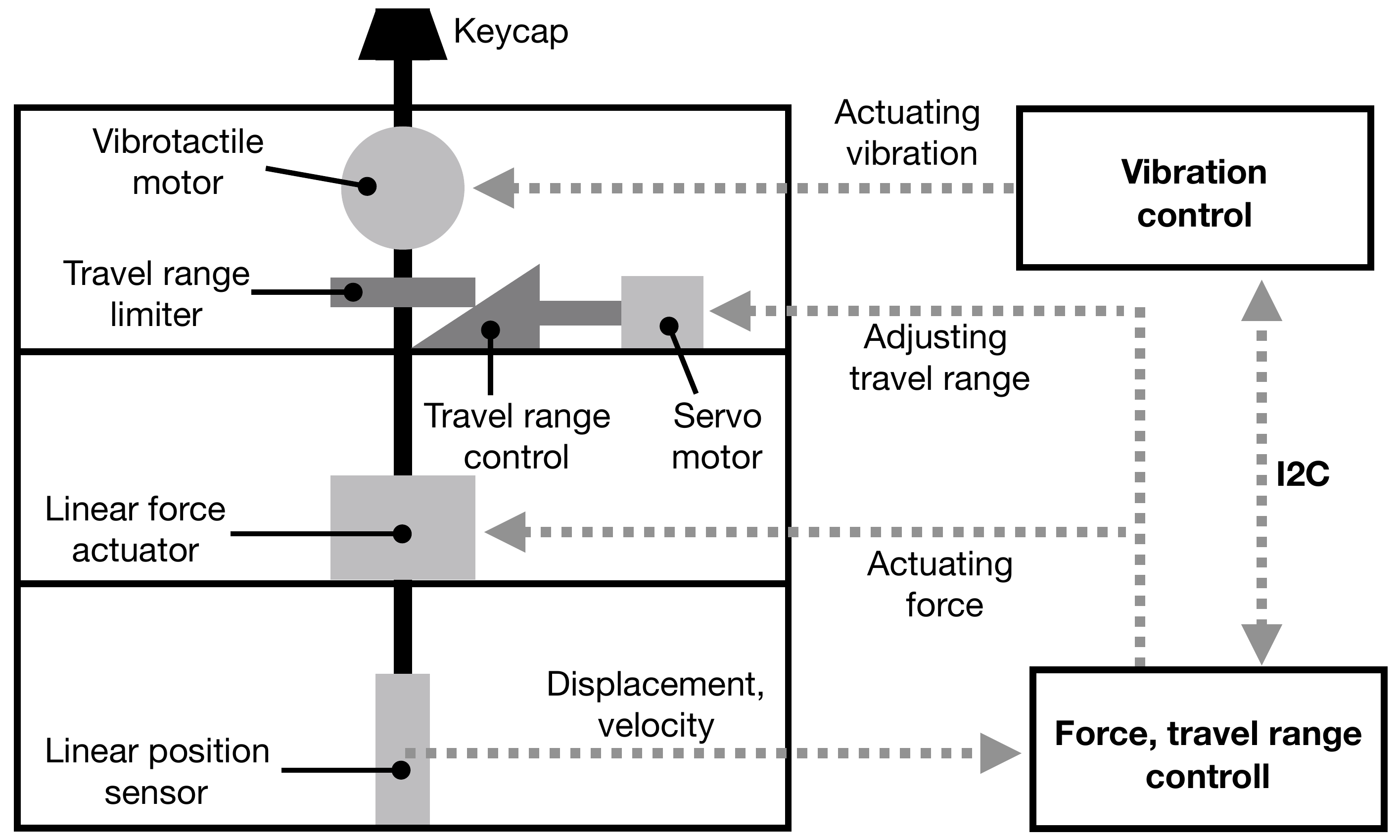}
  \caption{Physical simulator construction for haptic rendering of FDVV models. 
  Our simulator includes a 1D sensor that tracks displacement, a 1D force actuator delivering various levels of forces, and a servo motor drives the travel-range-adjustment component. The components are controlled by a microprocessor. 
  The other microprocessor controls a vibrotactile motor mounted near the keycap. 
  }~\label{fig:illustrated_system}
\end{figure}

Ours is the first physical simulator capable of the high-fidelity rendering of FDVV models. An overview is given in Figure \ref{fig:illustrated_system}.
Our first design goal was to provide the high-frequency response and high-resolution rendering of forces and vibrations typical of buttons.
The second was to enable full control from the software side. 

\textbf{Sensors and actuators:} 
Figure~\ref{fig:illustrated_system} presents the four main components: 
(1) a linear force actuator (Moticont HVCM-025-022-003-01), 
(2) a linear position sensor (LVDT MHR 250, \yc{resolution: 0.05 mm}), 
(3) a voice coil acting as a vibrotactile motor (Tectonic Teax13C02-8), and 
(4) a servo motor (Tower Pro Micro Servo, torque: 1.8 kg/cm). 
The force actuator, the sensor, and the servo motor are controlled by an Adafruit ItsyBitsy M0 board, which serves as the main processor of the prototype. 
The vibrotactile voice coil is driven by an Arduino Uno board and wave shield (Adafruit Wave Shield for Arduino Kit). 
These two boards are connected via the I2C protocol. 
When adjustments to the \textit{overall travel range} are required, 
the ItsyBitsy board sends a command to the servo motor to adjust the location of the \textit{Travel Range Control}, which further alters the lowest reachable displacement of the \textit{Travel Range Limiter} and produces varying travel.
When vibrotactile feedback is required, this board communicates with the Arduino Uno via I2C and asks it to drive the vibrotactile motor (voice coil) to present pre-recorded wave files.

\textbf{Microprocessor design:}
Before the simulation, the actuation signals (see ``Iterative Compensation'') are uploaded to the main microprocessor (Adafruit ItsyBitsy M0) and it automatically sets the button travel range. 
During a simulation, the linear sensor constantly sends the reading value to the microprocessor.
A moving-average filter (window size = 25) is applied here for denoising the reading from the position sensor. 
After the microprocessor has processed the values sent, 
it calculates the current displacement of the button and estimates the user's pressing velocity. 
Then, it determines the corresponding Pulse-width modulation (PWM) signal and sends it to the linear force actuator. 
At the displacement where vibration starts, 
the microprocessor sends a command to the Arduino Uno for emitting the vibration.
A high operating frequency is used (1 kHz) for the ItsyBitsy M0 board. 

\subsection{Spatial and temporal accuracy}
We measured the spatial accuracy of the simulator via a probing device consisting of a linear actuator and a probe attached to a force gauge \cite{Liao}. The probing velocity was 0.5 mm/s.
We used this device to profile a 4~mm Cherry MX Clear button,
checking the intended outputs of the simulator against the measurement results.
Over multiple repetitions, we learned that the simulator can reproduce the force responses very accurately, 
with only 1.44 cN mean error (SD 1.68 cN).
\yc{Time spent on each displacement sensing and computing actuation command is about 0.3 ms. 
From the command of rendering force to force generated takes less than 1 ms.
Latency from sending the vibration command to its actuation is about 7ms. We compensated this latency by emitting vibration 300 $\mu$m prior to the target starting point.}

\subsection{Simulation procedure}

Prior to simulation, 
the actuation signals obtained from iterative compensation (see later) are uploaded to the microprocessor. 
Near the beginning of the press -- i.e., at 0.5--1.0~mm of distance traveled -- the microprocessor calculates the pressing velocity for the press. 
At least three timestamped samples are needed.
From those samples, the simulator computes velocity and switches to the corresponding actuation specification. 
That specification is used to determine the resisting force and the vibration for the sensed level of displacement.
To simulate the vibration, we followed an event-based approach wherein vibration recording is initiated at the right displacement ~\cite{event-based},
creating a snap-like sensation.

\section{Iterative Compensation}

A key objective in our work is to separate the model from the simulator.
Any force actuator has its own transfer function in the play 
that must be canceled out if an FDVV curve is to be simulated correctly.
To our knowledge, no prior work on button simulation has considered this issue,
which may explain the lack of empirical evaluations of these simulators.
To address the issue, we introduce an iterative compensation method shown in Figure~\ref{fig:iterative_diagram}:
\begin{figure}[h!]
\centering
  \includegraphics[width=0.99\columnwidth]{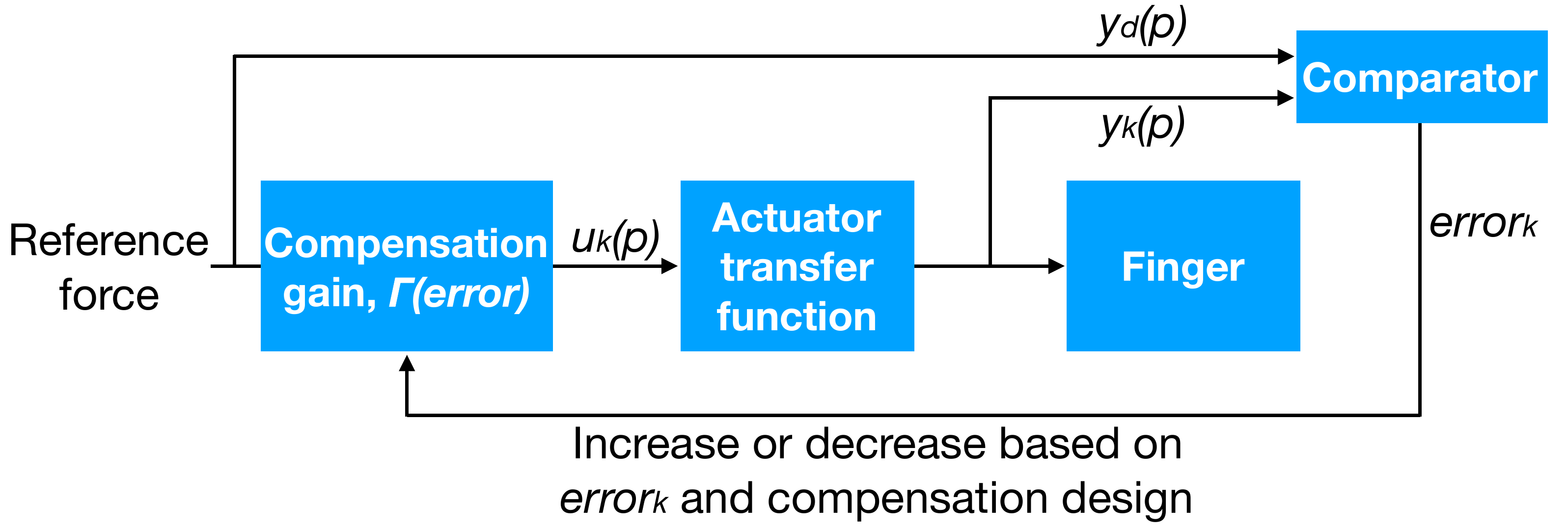}
  \caption{Iterative compensation finds a way to render an FDVV model on a given simulator plant.}
  ~\label{fig:iterative_diagram}
\end{figure}

The idea is to raise or lower the force-actuation signal amplitude of each displacement point until the desired resisting force is measured by the sensor against the keycap. 
Per-displacement and per-speed repetition can be applied 
until the desired FDVV signal is measured from the sensor. 
This iterative compensation process can be expressed as
\begin{equation}
u_{k+1}(p) = u_{k}(p) + \Gamma(error_{k}) (y_{d}(p)-y_{k}(p)), \quad p\in[1,n]
\end{equation}
Here, $u_{k}(p)$ is the actuation signal of a given displacement point $p$ in the current iteration, 
and $u_{k+1}(p)$ is the actuation in the next iteration signal of the same displacement point. 
$y_{k}(p)$ is the force detected from the sensor worn on the fingertip, 
and $y_{d}(p)$ is the desired target force at that given displacement point.
$\Gamma(error)$ represents the proportion of adjustment of the actuation signal that must be applied, based on the error value in the current iteration ($error_k$). 
The error from the current iteration is defined on a per-curve basis as follows:
\begin{equation} 
 error_k = \alpha\cdot\frac{\Sigma_{p=1}^{n}|{{y_d(p) -y_k(p)}|}}{n} + (1-\alpha)\cdot \max_{p\in [1,n]} |{y_d(p) -y_k(p)}|
\end{equation}
In this definition, two terms make up the error. 
The first is the overall difference between the target FD curve and the measured curve. 
The second is the displacement error at which the largest error is observed.
$\alpha$ is the weight applied between thees two, which we set to 0.7 based on experiences.  

\begin{figure}[t!]
\centering
  \includegraphics[width=0.99\columnwidth]{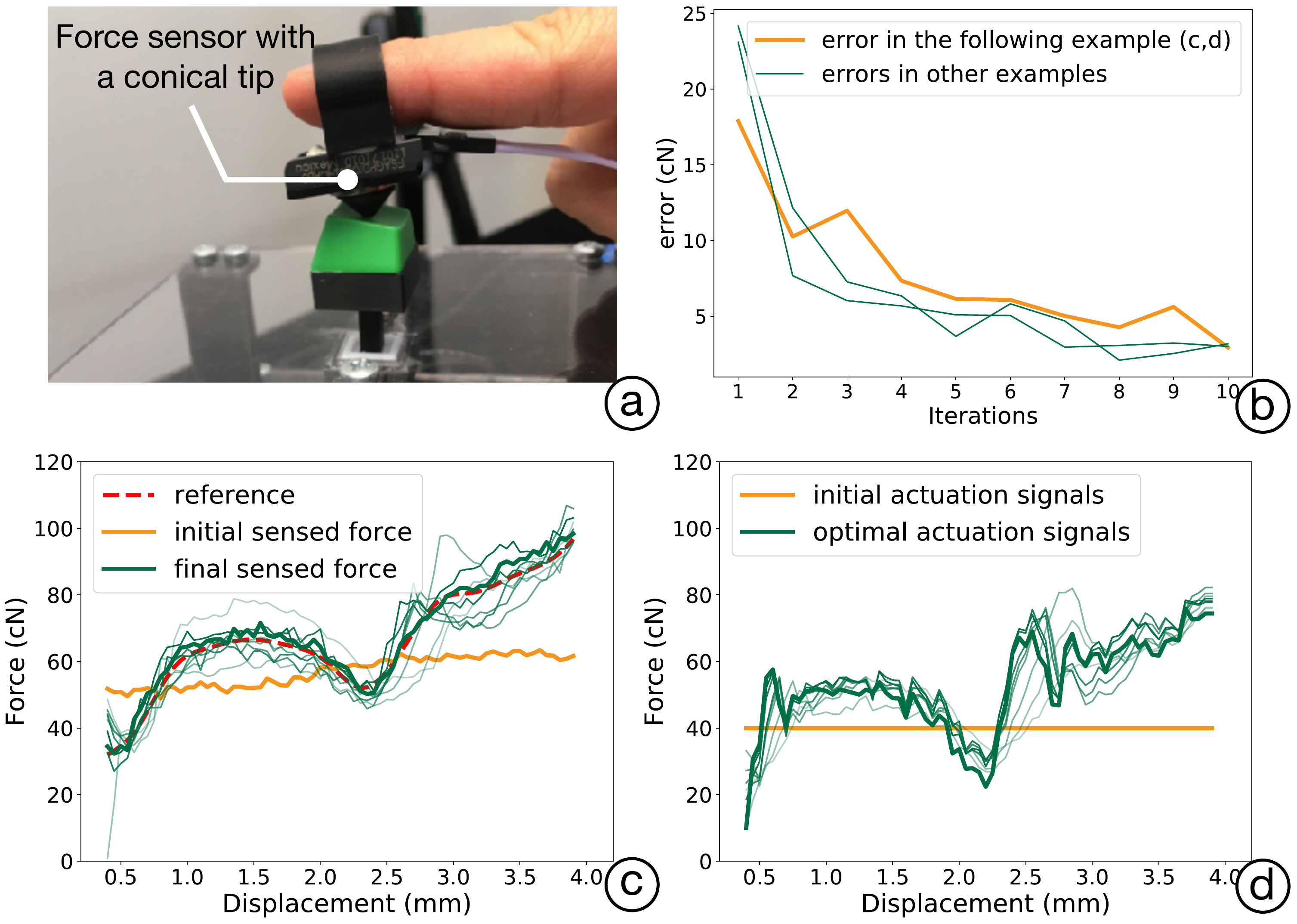}
  \caption{(a) During the iterative compensation, a force sensor is worn on the participant's fingertip which gathered force data and sent it to the controller. 
  (b) Some instances of the evolution of error values during the process. The blue curve represents the errors of (c,d) which fell from 17.89 cN in the first iteration to 2.93 cN in the 10th. 
  (c) An example keypress from which we can see that the sensor worn on the fingertip shows convergence with the reference after the compensation process is complete. (d) The actuation signals starting at a random force level and being gradually tuned. 
  Note that we transform the actuation signals linearly into force level (cN) that can be measured in a steady machine-probing situation.
  }~\label{fig:iterative_learning}
\end{figure}

\textbf{Procedure:} 
We first upload the reference data (FDVV model) to the simulator. 
A participant is asked to press the button repetitively in line with the speed indicator presented by a GUI. Figure~\ref{fig:iterative_learning} illustrates the whole process.
The force sensor is connected to the same microprocessor that runs the simulator. 
The sensor (response rate: 1 kHz) captures the force responses during a press by the fingertip. 
The value sensed is passed through a resistor--capacitor filter with a 498 Hz cutoff rate, to reduce within sensor noise.
All the resisting force samples within a 50~$\mu$m interval are aggregated and averaged for this displacement point.
In our experience, the procedure converges after only 8--12 presses;
however, each button needs to be modeled at multiple velocities.
With four velocities, there are 240 presses in total (4 velocities $\times$ 4 rounds $\times$ 15 presses) to compute its actuation signals.
A typical example is presented in Figure~\ref{fig:iterative_learning} (b): 
$error_k$ decreases rapidly to below 3~cN within 10 presses.

\begin{figure}[t!]
\centering
  \includegraphics[width=0.99\columnwidth]{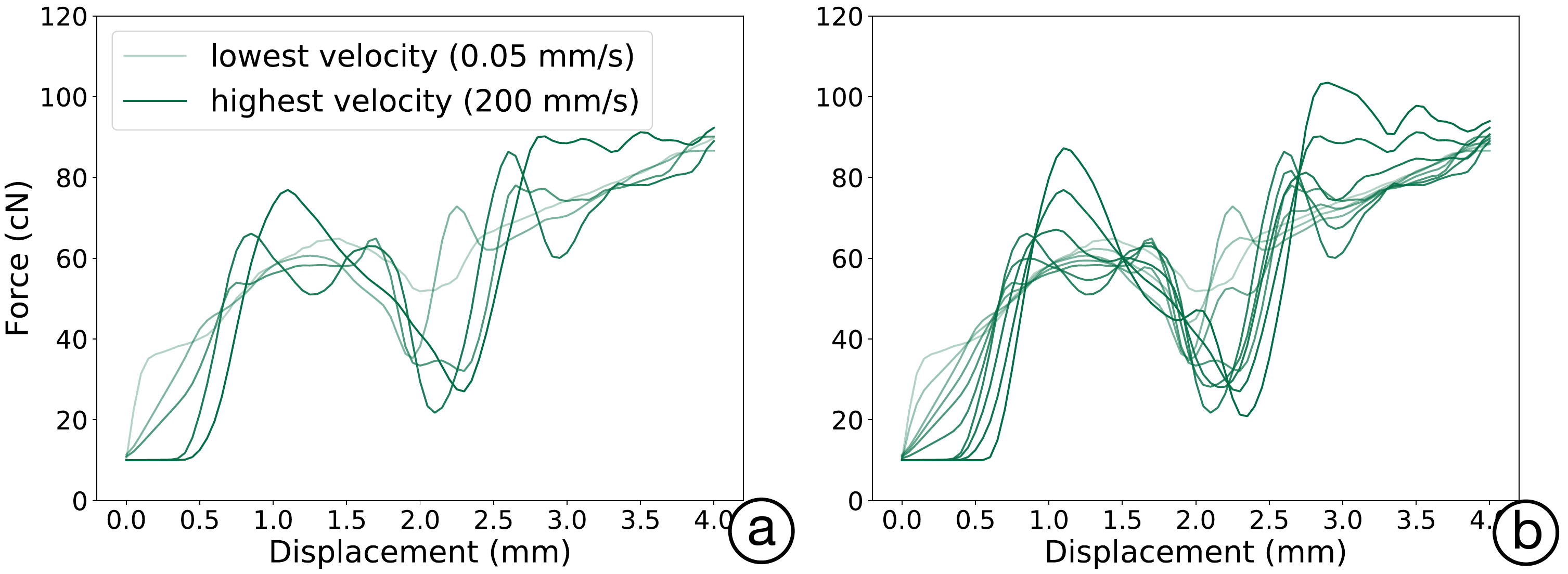}
  \caption{Final, linearly interpolated actuation signals: 
  (a) Example actuation signals before linear interpolation, here shown for a Cherry MX Clear with various velocities, which are the same as in Button Capture.
  (b) The same data after linear interpolation. }~\label{fig:interpolated_signals}  
\end{figure}

\textbf{Outputs:} After the process is complete, 
the microprocessor records the actuation signals that resulted in the minimal error.
For a given reference force curve, 
we ran the iterative compensation process four times, obtaining four series of actuation signals. We then averaged these at each displacement and finally applied a Gaussian filter ($\sigma=1.2~mm$) to smooth the signals.
After all the force-actuation signal curves were obtained (see Figure~\ref{fig:interpolated_signals}~(a)), we linearly interpolated the signal curves to form denser, more continuous curve sets that responded to more velocity changes (see Figure~\ref{fig:interpolated_signals}~(b)).  

\subsection{Optional human-in-the-loop vibration tuning}

As described earlier, sometimes the vibration emitted at the snap point is weak and our sensor does not reliably pick it up accurately. 
Also, sometimes when vibration is measured as a soundwave, it may fail to reproduce the same sensation when reproduced using the vibration motor.
In order to produce the desired snap sensation, the vibration needs to be accentuated. 
To this end, we devised a human-in-the-loop method for tuning the vibration response at the simulator side.
\yc{To further render more realistic vibrotactile feedback, future work should consider more sophisticated modeling and rendering techniques\cite{Wellman1995,951362,7744626,physvib}.}

\begin{figure}[t!]
\centering
  \includegraphics[width=0.99\columnwidth]{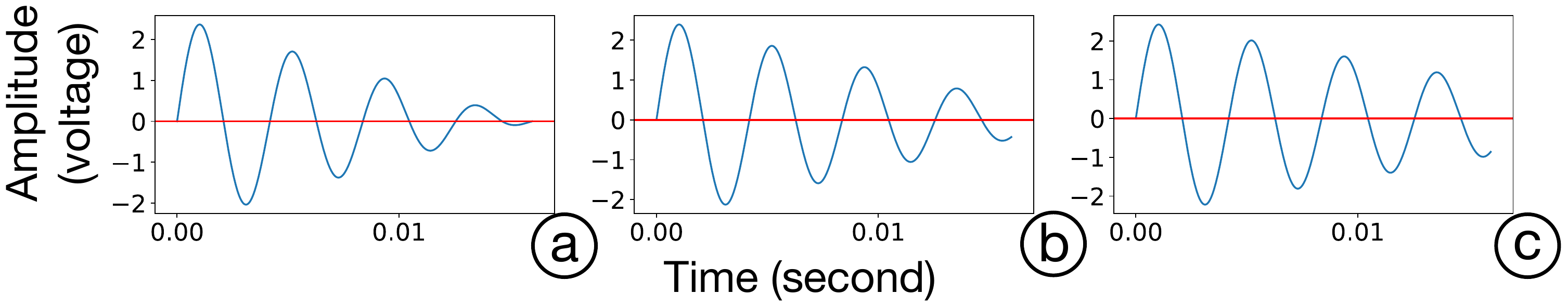}
  \caption{\yc{Example decaying sinusoidal-wave templates. They share the same frequency (239 Hz) and duration (16 ms) which are captured from Cherry MX Clear, and the amplitudes and shapes vary for tuning: (a) amplitude decreases from $\pm 2.43$ to $0$ Vol, (b) amplitude decreases from $\pm 2.43$ to $\pm 0.3$ Vol, and (c) amplitude decreases from $\pm 2.43$ to $\pm 0.6$ Vol.}}~\label{fig:vibration_matching}
\end{figure}

\textbf{Procedure:} 
\yc{We obtained 3 features from vibration measurements: (1) vibration onset, (2) duration and (3) frequency.
Afterward, an algorithm generated several vibration templates that match the recorded vibration and duration. 
The generated templates are decaying sinusoidal waves with various frequencies and accentuated amplitudes. The generative method we follow is by Park \textit{et al.}\cite{sinusoidal}.}
\yc{These sound-wave templates were uploaded to the Arduino Uno, which simulates them, using the vibrotactile motor (see Figure~\ref{fig:illustrated_system}).
Some templates are shown in Figure \ref{fig:vibration_matching}. 
Finally, as part of our human-in-the-loop tuning process, we asked a human observer to press the simulated button at the pace shown in the animation (see above). 
The user rated each button-design--vibration combination. We repeated this process for all velocities. The best-rated vibration sets were selected as the final actuation signals for that button.}

\section{A User Study: Perceived Realism}

We assessed the perceived realism of the rendered buttons in a controlled study.
We adopted the idea of ABX test as used in psychophysics for comparing two sensory stimulus options for identifying a target ~\cite{sensory_evaluation,clark1982_abx,virtual,Munsonwa}.
A participant tries a real reference button (X) and is then asked to press two simulated buttons (A, B) and decide which offers a more realistic rendering of it. 
The A, B buttons are rendered via the same physical simulator,
and the user can try out the three buttons as many times as desired.
We compared our FDVV models against speed-agnostic FD models
because it represents the prevailing understanding of button tactility represented in academic literature, hobbyist groups, and manufacturer datasheets. 

\subsection{Method}

\textbf{Participants: }
We recruited 12 participants (6 females) from a local university, of ages 21--41 (mean 29.75). 
All of them reported typing and other button-pressing experience. 
They were rewarded with a movie ticket (valued at 14 euros) for the 60-minute study.

\textbf{Task and apparatus: }
The study compared six physical buttons listed previously in ``Button Capture''.
They differ in characteristics, but all are realistic.
All the buttons were captured and transformed into (1) a single-FD model and (2) FDVV models.
Actuation signals were computed as described in the previous section.

To prevent users' haptic judgments from being biased by their vision, the simulator was placed inside a black box with a hole, into which the user reached to press. 
The target buttons and the simulator were at the same height. 
The participants were asked to wear a headset playing white noise and earmuffs, to isolate hearing during the study.
A graphical interface showed which of the two buttons (labeled A and B) was currently active.
The information displayed was the name of the target button (one of the six), the simulated button's label (A or B), and the current trial number. 
Double-blind administration was employed: 
neither the experimenter nor the participant knew which button (A vs. B) used FDVV and which used FD.

\textbf{Procedure and experiment design: }
The participants were told about the simulator and the purposes of the study.
They were asked to explore the real buttons once, with different pressing velocities. 
We did not repeat this instruction during the study proper, though; we let them decide what was natural for them. 

\begin{figure}[t!]
\centering
  \includegraphics[width=0.99\columnwidth]{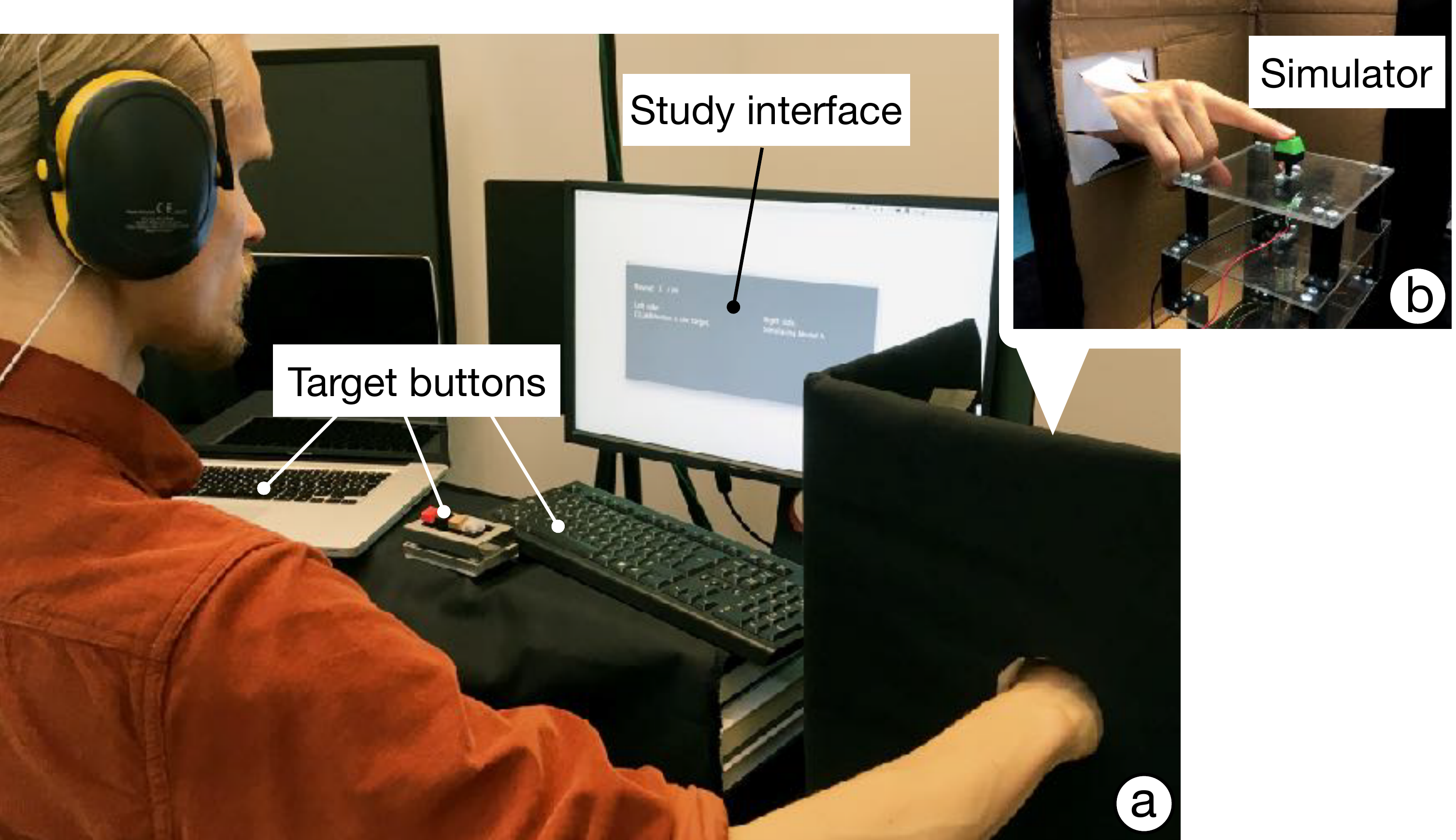}
  \caption{We assessed the perceived realism of the buttons in an ABX task. Firstly, (a) a participant was shown a (real) reference button (X). 
  Then (b) the participant could try one simulated button (either FD- or FDVV-based) on the simulator at a time. 
  The display showed the label for the active button (A or B). Afterward, the participant indicated which of the two (A or B) was more realistically rendering X.
  }~\label{fig:study_setup}
\end{figure}

Again, each round featured a \emph{reference button} and \emph{two simulated buttons}.
The interface identified the reference button and the label of the currently active button   (see Figure~\ref{fig:study_setup}).
The participants were told that there are two simulated buttons in each round, denoted as \emph{button A} and \emph{button B}. 
In each round, participants were instructed to press the reference button and to feel it.
After that, they were asked to try the alternative simulations, labeled A and B.
They could switch among the three buttons as many times as they wished.
When ready to make their judgment, they were asked
to indicate which button had more realism (A, B, or equal) and to rate the perceived realism of A and B separately, on a seven-point Likert scale. 
After the study, an interview was conducted. 
We gave a three-minute break after every 20 minutes of button presses, to minimize fatigue.
There were six rounds for each of the six buttons, making 36 trials in total.
The trial order of the six button designs was counter-balanced by Latin square.
The assignment of the FDVV and FD models to the labels, A and B, was randomized at each trial.

\subsection{Results}

\begin{figure}[b!]
\centering
  \includegraphics[width=0.99\columnwidth]{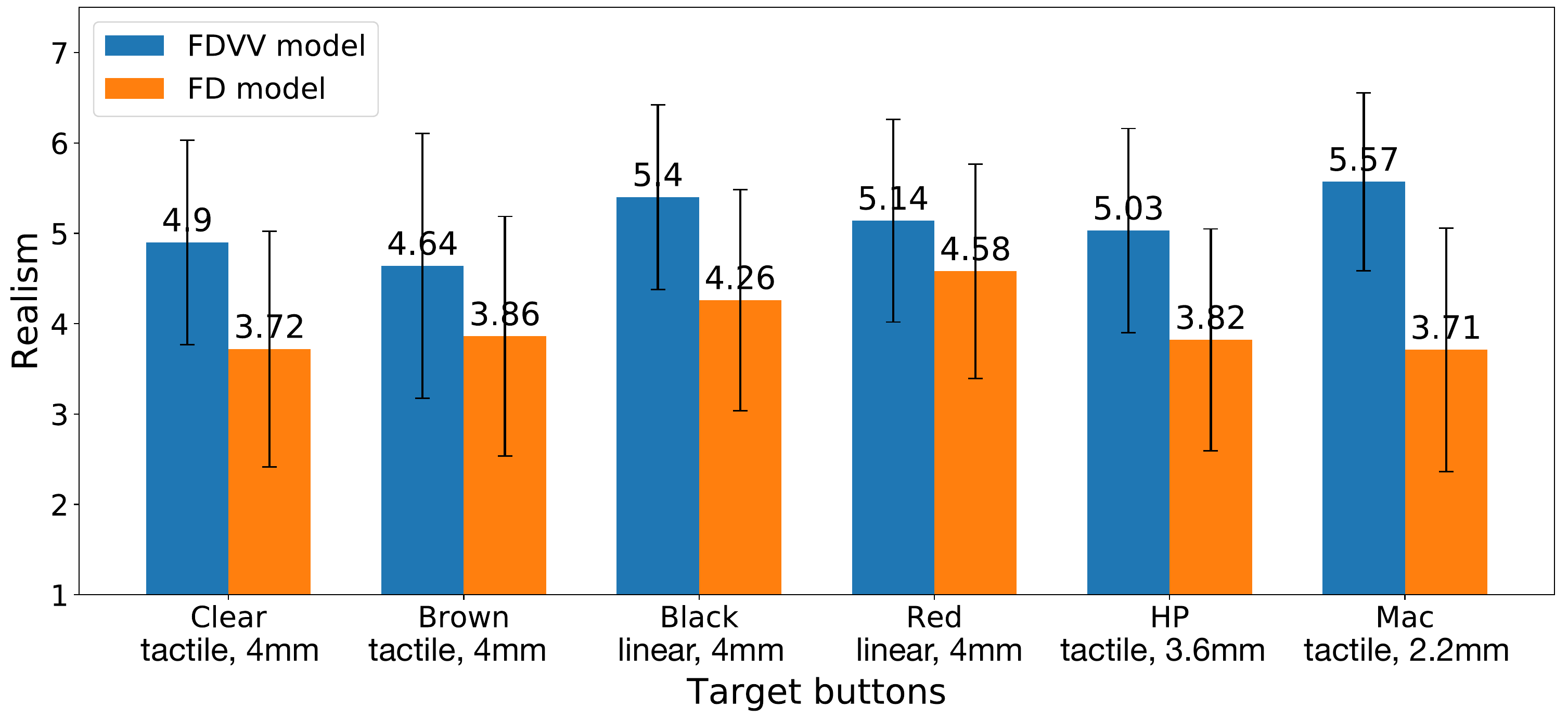}
  \caption{Users in the ABX identity-matching study rated FDVV-based simulations as more realistic than FD-based simulations. 
  Statistically significant differences were found for all the target buttons. The error bar in the figure is 1 STD.}~\label{fig:study_results}
\end{figure}

The results support the FDVV approach.
It was associated with higher perceived realism for all the simulated buttons. 
The participants chose the FDVV model as more realistic 77.31\% of the time. 
An overview is shown in Figure~\ref{fig:study_results}.
We examined the ratings further by using Wilcoxon Signed Ranks Tests. 
The analysis showed that there are statistically significant differences for each target button between the FDVV and single-FD models:
\begin{enumerate}\compresslist
\item For the \textit{Clear} button, the median FDVV model ranks (mdn 5.16, mean 4.9, STD 0.72) were significantly higher than the median single-FD ones (mdn 3.83, mean 3.72, STD 0.92), with $Z = -3.06$, $P = 0.002$.  
\item For the \textit{Brown} button, the median FDVV model ranks (mdn 4.92, mean 4.64, STD 1.02) were significantly higher than the median single-FD ones (mdn 3.83, mean 3.86, STD 0.87), with $Z = -2.748$, $P = 0.006$.
\item For the \textit{Black} button, the median FDVV model ranks (mdn 5.33, mean 5.4, STD 0.53) were significantly higher than the median single-FD ones (mdn 3.83, mean 4.25, STD 0.72), with $Z = -2.94$, $P = 0.003$.
\item For the \textit{Red} button, the median FDVV model ranks (mdn 5.33, mean 5.14, STD 0.9) were significantly higher than the median single-FD ones (mdn 4.67, mean 4.58, STD 0.54), with $Z = -2.158$, $P = 0.031$.
\item For the \textit{HP keyboard} (spacebar), the median FDVV model ranks (mdn 5.33, mean 5.03, STD 0.58) were significantly higher than the median single-FD ranks (mdn 3.83, mean 4.0, STD 0.85), with $Z = -2.987$, $P = 0.003$.
\item For the \textit{MacBook Pro keyboard} (spacebar), the median FDVV model ranks (mdn 5.42, mean 5.57, STD 0.57) were significantly higher than the median single-FD ranks (mdn 4.67, mean 3.67, STD 0.9), with $Z = -3.06$, $P = 0.002$.
\end{enumerate}

From Figure \ref{fig:study_results}, we see that the smallest difference between the FDVV and FD simulation was for the Red button.
This is a linear button, and the FD model offers a reasonable simulation.
One participant stated, \textit{``The red one has a smooth (linear) feeling, and it's lighter than the other buttons. 
I feel two models with a similarly smooth and light feeling, so it's hard to tell the differences.''
}

\section{Applications}

We show three applications exploiting the approach.

\subsection{Human-in-the-loop button optimization}

\begin{figure}
\centering
  \includegraphics[width=0.99\columnwidth]{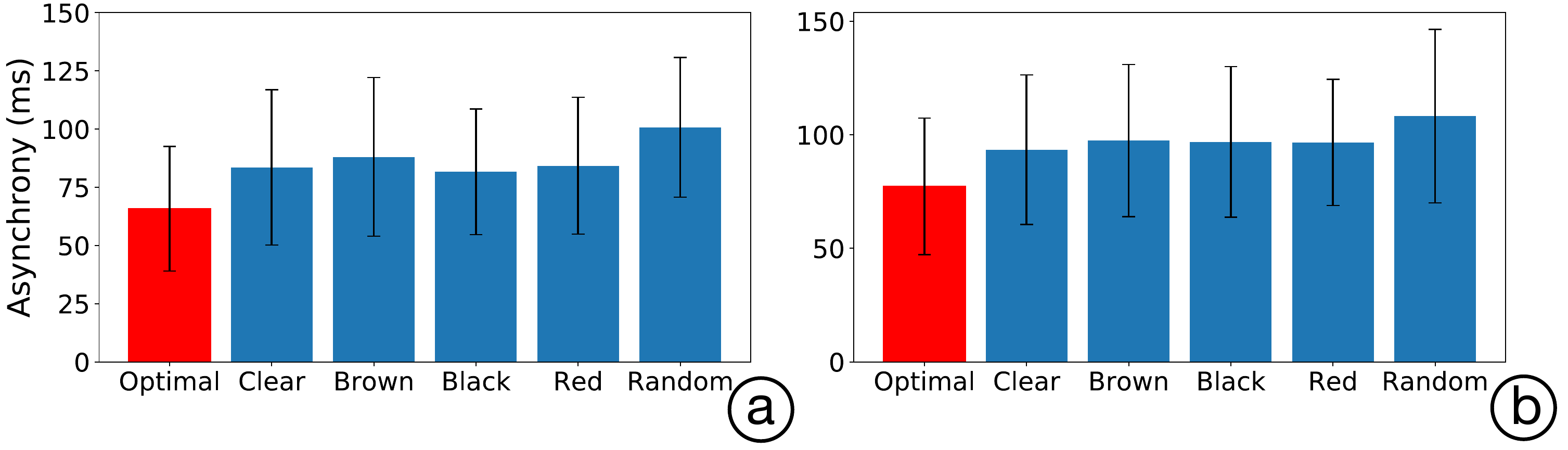}
  \caption{Button optimization significantly decreased mean asynchrony in temporal pointing for both (a) easy and (b) hard tasks. The error bar in the figure is 1 STD.}~\label{fig:optimization_results}
\end{figure}

Firstly, we demonstrate the optimization of a button for an interactive task. 
We look at a temporal pointing task with no visual feedback~\cite{modelling_errorrate}.
It resembles games where a response must be elicited at just the right time (for instance, to catch an enemy). 
Our goal is to optimize the button's FDV design and also its activation point. \yc{Velocity-dependent properties (the last V in FDVV) were excluded due to assuming a person presses a button at a similar speed all the time.} 
The optimization we used was Bayesian optimization (BO),
which is favorable for conditions wherein evaluations are noisy and expensive \cite{shahriari2015taking}.  
The objective in BO is to minimize a user's mean asynchrony~\cite{modelling_errorrate,Repp2005},
or the mean difference in time between the target and the user's elicited response.
Figure \ref{fig:optimization_results} depicts some example FDV models generated by the optimizer.
To keep the study below one hour in total length,
we limited the BO's task to three control points and two other button parameters.
We mapped this to the force-actuation signals,
which the BO then manipulated
so that once a new design is sent to the simulator, users can instantly try it out without iterative learning. 

\textbf{Method:}
We recruited 10 participants (4 females) from a local university, of ages 20--40 (mean 26). 
The study had two phases, \emph{training} and \emph{validation}.
In the training phase, 
they were asked to press the button when the LED strip showed a bullet to have reached the center of the target zone.
Two levels of task difficulty (\emph{easy} = 100 pixels/second; \emph{difficult} = 150 pixels/second) were used.
For each level, 27 trials were collected and used to compute the mean asynchrony score.
The whole process took about 60 minutes. 
Short breaks were given after every 15 minutes of presses.

Our BO implementation was based on Python's GPyOpt library\footnote{~See http://sheffieldml.github.io/GPyOpt/}. 
In each iteration, it changed the parameter values of the button model.
We had three control points in the FDV and three other parameters: $x1, x2, x3$ (displacements of three control points) and $y1, y2, y3$ (actuation signals of those control points). 
The ranges of these six parameters were $x1 \in [0,1), x2\in[1,3), x3\in[3,6.2)$ and $y1, y2, y3 \in [20,180]$. 
The other two additional parameters were activation point, $p_a$, and vibration point, $p_v$. 
The ranges of these two parameters were $p_a, p_v \in [0.5,5.5]$.
An additional microprocessor, an Arduino Uno, was set to drive an LED strip (Adafruit DotStar) that displayed an array of LED lights with the bullet animation. 
This microprocessor was connected to the button simulator by software serial port.
When the button passed the activation point, the microprocessor of the simulator sent a triggering signal to the LED strip, 
which would then calculate the temporal error of this press.
After 20 presses, we calculated the mean asynchrony and sent the information back to the simulator. 
Then, the optimizer created a new design of button for the next iteration, based on the data collected, and triggered the simulator to reconfigure itself accordingly.

In the testing phase, a week after the training phase, 
the optimized button was compared to other non-optimized buttons: four 4~mm mechanical ones (Cherry MX Clear, Brown, Black, and Red) and a random button design \yc{(all parameters are randomly given within the defined range)}.  
\yc{Two difficulty levels are assigned in a counter-balanced order.}
Each button appeared twice per difficulty level, in counter-balanced order, 
and the participant needed to press the button 20 times in every trial.
In total, we collected 240 observations of mean asynchrony (10 participants $\times$ 2 difficulty levels $\times$ 2 rounds $\times$ 6 buttons).

\begin{figure}[t!]
\centering
  \includegraphics[width=0.99\columnwidth]{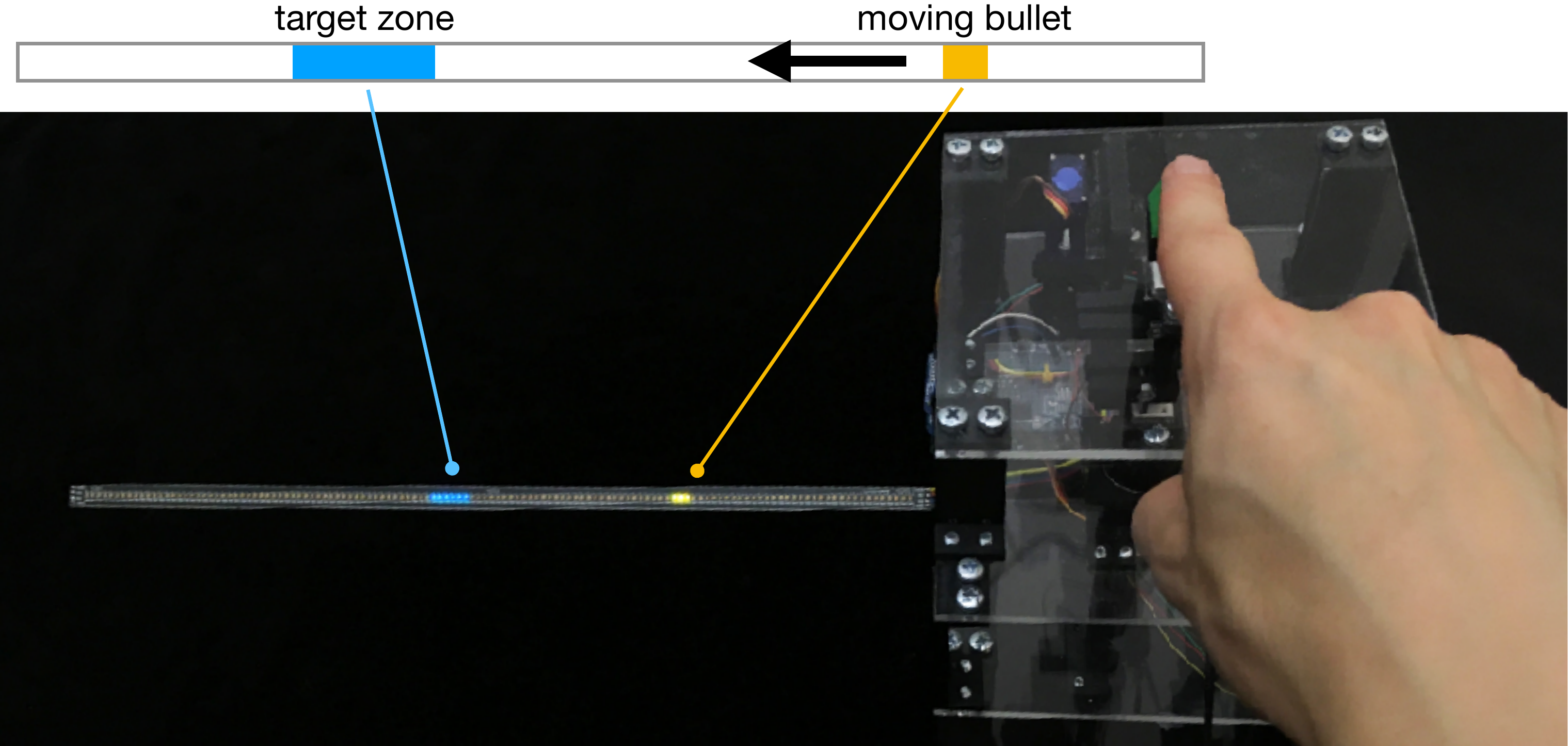}
  \caption{Button optimization was evaluated in a temporal pointing task. Users were asked to hit a temporal target by pressing a button at the right time, as a moving bullet reaches the target zone. After the optimal design was learned, the actuation signals were translated back into an FDV model via the capturing process.}~\label{fig:optimization_setup}
\end{figure}

\begin{figure}[t!]
\centering
  \includegraphics[width=0.99\columnwidth]{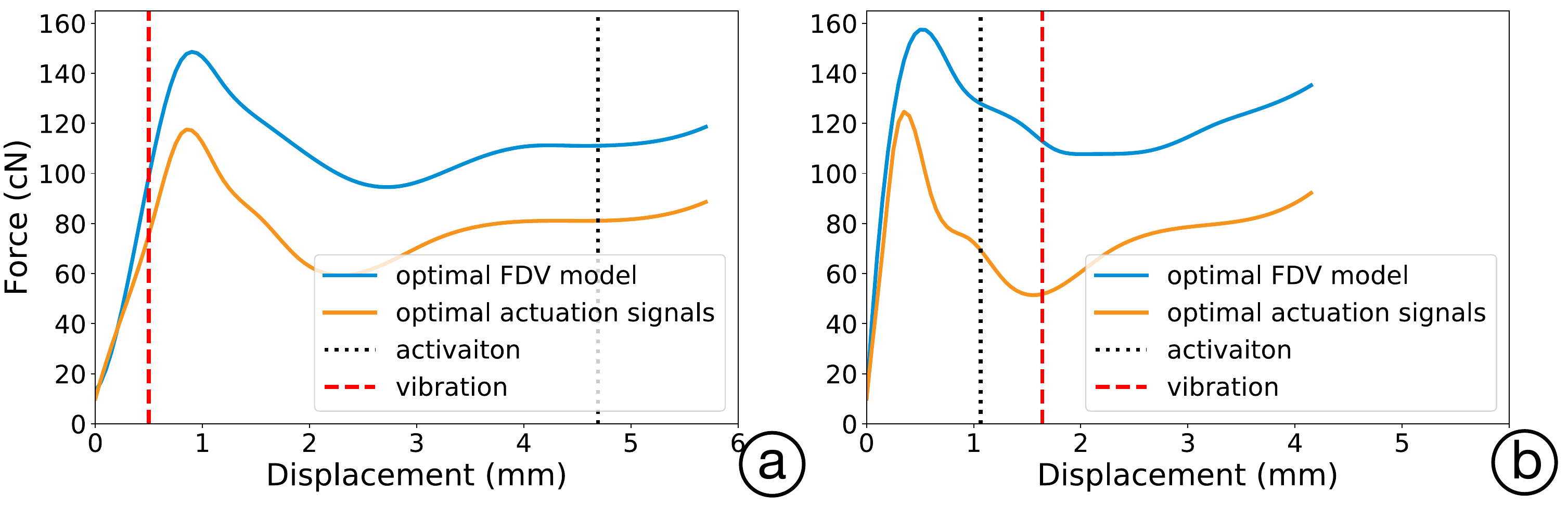}
  \caption{An example outcome of Bayesian optimization for a user under the (a) ``Easy'' and (b) ``Hard'' task condition.}~\label{fig:optimization_ex_outcome}
\end{figure}

\textbf{Results:}
In Figure~\ref{fig:optimization_ex_outcome}, we present some instances of optimal button designs. 
For the difficulty level \textbf{Easy}, the resulting mean asynchronies were 65.8 ms (STD 6.15), 83.6 (STD 7.65), 88.04 (STD 7.82), 81.65 (STD 6.19), 84.28 (STD 6.75), and 100.78 (STD 6.89), for the optimal, clear, brown, black, red, and random button, respectively. 
For the \textbf{Hard} difficulty level, the resulting mean asynchrony values are 77.3 ms (STD 6.89), 93.43 (STD 7.56), 97.48 (STD 7.69), 96.9 (STD 7.61), 96.65 (STD 6.36), and 108.22 (STD 8.77), for the optimal, clear, brown, black, red, and random button, respectively. 
\yc{A two-way repeated measures ANOVA was conducted. The main effect of buttons on mean asynchronies is significant, $F(5,95)=10.724$, $p < 0.001$. The \emph{post-hoc} tests with Bonferroni correction confirmed the optimal button design as indeed outperforming the rest ($p<0.05$).}

\subsection{Interactive button design} 

\begin{figure}[t!]
\centering
  \includegraphics[width=0.99\columnwidth]{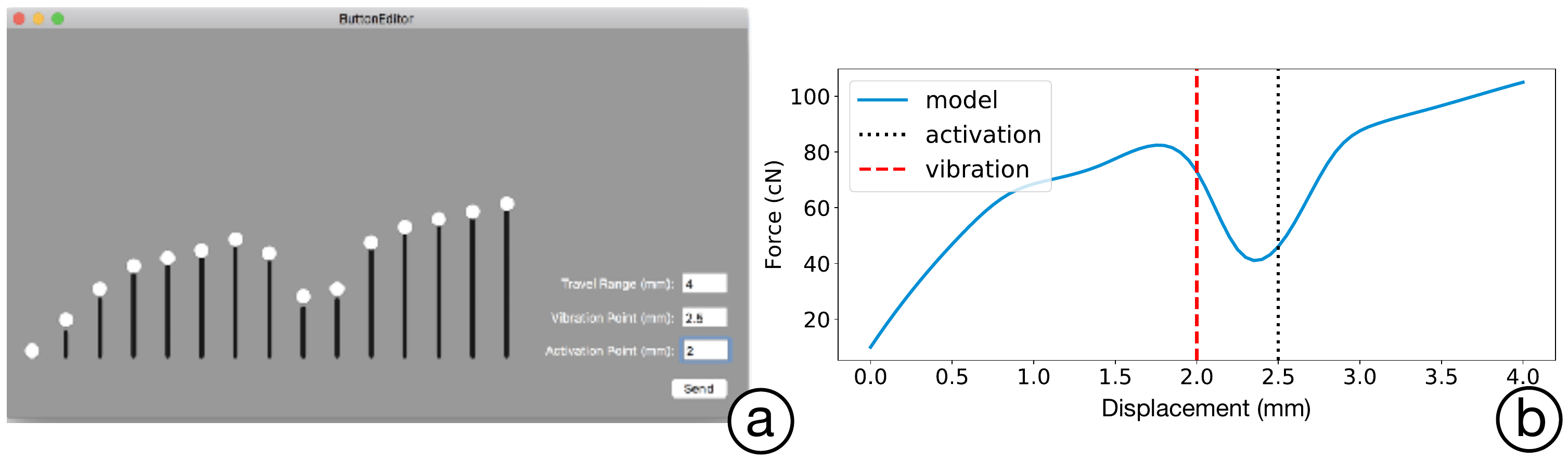}
  \caption{The button-editing tool: (a) The design tool allows the designer to freely create and edit low-parametric FDV models. 
  (B) An edited model can be processed and then simulated via the simulator.}~\label{fig:button_editing} 
\end{figure}

We implement an \emph{interactive button-editing tool} that lets engineers freely create and edit button designs (\autoref{fig:button_editing}).
Using the tool (implemented on macOS using Swift), designers can manipulate force levels through 15 draggable control points to create a single-FD curve.  
Travel range, activation point, and vibration point can be edited textually.
Next, the model is converted to high-dimensional force actuation signals (with B-spline fitting and iterative compensation).
Finally, these signals are used to simulate the button.
Several button designs implemented by the interactive button-editing tools are presented below, under ``Innovative button designs.'' 
While the tool allows for editing of one FDV model at a time, our backend (Python) also allows for more advanced editing, such as assigning a specific FD curve segment to be pressed or released, and merging multiple FD curves into an FDVV model.

\subsection{Prototyping Innovative Buttons}

Our setup enables prototyping innovative and extraordinary buttons that go beyond commodity designs. 
We give five examples here. 
The first two are instances of FDVV models, 
while the remaining are buttons that can be run on the simulator but can not be expressed as FDVV models.

\begin{figure}[t!]
\centering
  \includegraphics[width=0.99\columnwidth]{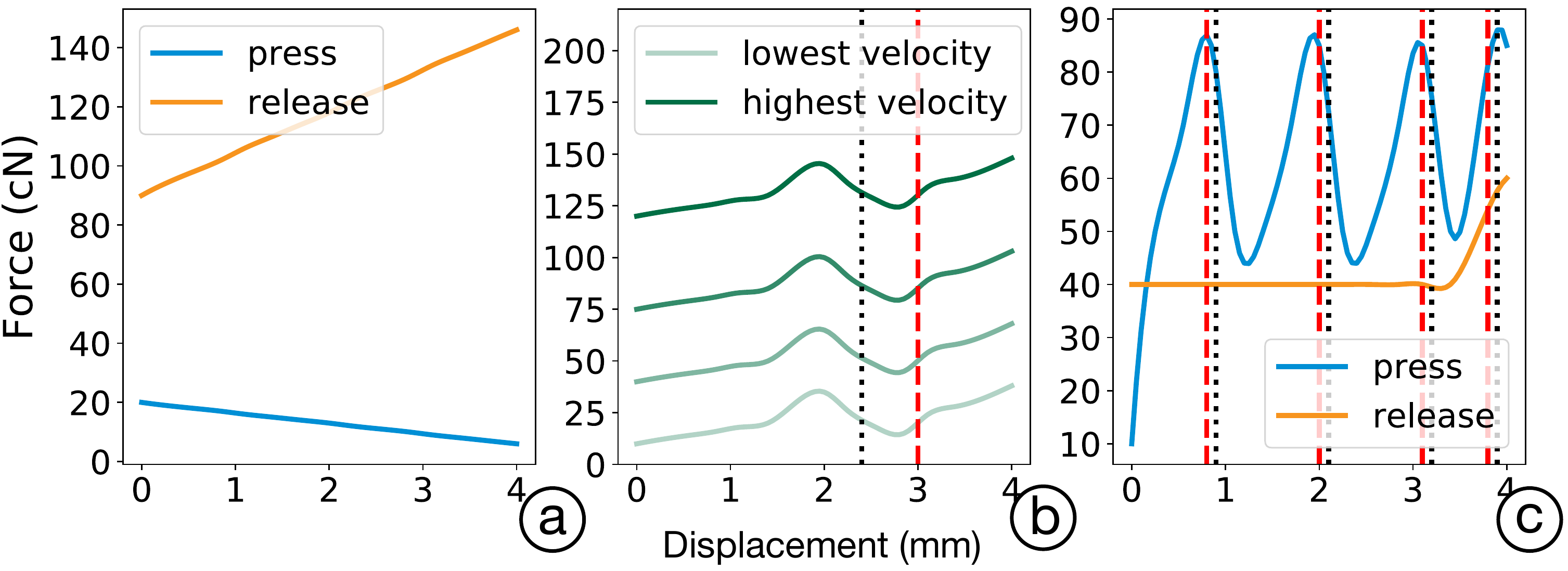}
  \caption{Three innovative button designs realized with our setup: 
  (a) a button for very fast tapping, 
  (b) a non-Newtonian-fluid button, and
  (c) a multi-level button.}
  \label{fig:innovative_buttons}
\end{figure}

\textit{1. A fast tapping button}: 
While humans can only reach about four presses per second in tapping tasks~\cite{Repp2005}, 
we can increase this capacity via a novel button design. 
The principle is this: once a press is detected, the button drops to the bottom and returns automatically. 
This can be especially useful for content requiring high-frequency rhythmic tapping,  
such as music games (see Figure~\ref{fig:innovative_buttons}~(a)).
One of the authors reached 8 presses per second with this button.

\textit{2. A non-Newtonian-fluid button}: In non-Newtonian fluids, viscosity changes under force: the medium becomes either more liquid or more solid. We can design a button that adjusts its stiffness following pressing velocity; see Figure~\ref{fig:innovative_buttons}~(b).
As a result, the button is softer when being pressed gently, but the resisting force increases during fast pressing. 
This design can be used to prevent accidental touches.

\textit{3. A multi-level button}: 
We can extend the input modality of a button by giving multi-level haptic feedback. 
For example, a typical 4~mm button can be divided into several depth ranges. 
Through the provision of distinguishable haptic signals, a user can effectively activate different functions of the button by pressing down to different depths; see Figure~\ref{fig:innovative_buttons}~(c).
This can be useful for easier use of single-button devices (e.g., in tuning the luminance of a lamp).

\textit{4. Vibration cues}: We can deliver rich temporal information through continuous vibrotactile cues while the press is at the bottom. 
This interaction can enhance the effectiveness and efficiency of dwell-press applications\cite{LiaoDwell}. For instance, when the shutter button of a camera is pressed and the camera is continuously shooting, vibration ticks can help the user easily count the number of shots via humans' haptic sense.

\textit{5. A dynamically returning button}: 
In certain situations, it might be desirable to avoid fast repetition with a given button. In the example of fighting games or shooting games, many attacks involve a cooldown time -- i.e., a minimum duration before the next use of the relevant ability. Our simulator can easily render buttons with just such dynamic returning time, as demanded by the game content.

\section{Conclusion}

We have shown that the FDVV approach proposed as an extension to the dominant FD model increases the scope and perceived realism of button simulations. 
While this extension was motivated by prior literature,
several engineering problems were solved to capture and simulate FDVV models.
The added complexity notwithstanding, the core model is understandable and offers practitioners a workable starting point.
We have demonstrated the benefit of model--simulator separation via three applications: human-in-the-loop button optimization, interactive button design, and examples of innovative buttons that would be tricky to realize without this approach.

While the FDVV approach permits greater realism, and opens many new practical possibilities, 
our results point also to clear opportunities for further improving realism. 
Firstly, the iterative compensation method can be enhanced,
to better capture the \emph{dynamic} effects of damping on the actuation signals.
Our approach was to try to cancel the effect of the transfer function,
but future research could consider learning a data-driven model of the black box through machine learning. 
This should be coupled with a clever controller design.
Secondly, structural vibration can be modeled better.
We opted to measure the snap-like vibration as a sound wave,
but sometimes this was inadequate to reproduce the felt sensation of the tactile bump.
To enhance the realism of the snap point, 
researchers could consider applying better measurement, modeling and controlling methods for handling the vibration\cite{951362,physvib}.
This would eliminate the human-dependent part of the vibration modeling.
While these changes would increase realism,
perfect button simulation entails simulating the texture and shape of the keycap, the sound emitted, etc.
Finally,
once a button has been designed and tested, 
a suitable electromechanical design should be fabricated.
This presents an interesting challenge for future work aimed at bridging the gap between buttons \emph{in vitro} and buttons \emph{in vivo}.  

\section{Open Science}
The materials and data in this paper are released on our project page at \url{http://userinterfaces.aalto.fi/button-design}. The materials include 3D models, circuit design, component specifications, construction details of the simulator, and the programs for controllers. 
\yc{The cost for the simulator construction is about \$550.
From capturing a button to simulation via our pipeline takes about an hour (5 minutes for capturing, 30 minutes for data preprocessing and FDVV modeling, and 30 minutes for iterative compensation and simulation).}
All the data for button measurements and experiments are released too. The materials also complement Figure \ref{fig:fdvmodel} and Figure \ref{fig:interpolated_signals} with the graphs of other buttons.

\section{Acknowledgement}
This work has been funded by the European Research Council (ERC) under the European Union’s Horizon 2020 research and innovation programme (grant agreement No 637991) and by Korea Creative Content Agency (grant agreement No  R2019020010).

% BALANCE COLUMNS
\balance{}

% REFERENCES FORMAT
% References must be the same font size as other body text.
\bibliographystyle{SIGCHI-Reference-Format}
\bibliography{proceedings}

\end{document}